\documentclass[final,1p,times]{elsarticle} 
\usepackage{graphicx} 
\usepackage{amssymb} 
\usepackage{amsthm} 
\usepackage{lineno} 

\journal{Nuclear Physics A} 
\begin{document} 

\begin{frontmatter} 


\title{Viscosities in the Gluon-Plasma within a Quasiparticle Model}

\author{M.\ Bluhm$^{a,b}$, B.\ K\"ampfer$^{a,c}$, K.\ Redlich$^{d}$}

\address[a]{Forschungszentrum Dresden-Rossendorf, 
Institut f\"ur Strahlenphysik, PF 510119, 01314 Dresden, Germany}
\address[b]{CERN, Physics Department, Theory Devision, 
CH-1211 Geneva 23, Switzerland}
\address[c]{Technische Universit\"at Dresden, 
Institut f\"ur Theoretische Physik, 01062 Dresden, Germany}
\address[d]{University of Wroclaw, 
Institute of Theoretical Physics, PL-50204 Wroclaw, Poland}

\begin{abstract} 
A phenomenological quasiparticle model, featuring dynamically generated 
self-energies of excitation modes, successfully describes lattice 
QCD results relevant for the QCD equation of 
state and related quantities both at zero and non-zero net baryon density. 
Here, this model is extended to study bulk and shear viscosities of the 
gluon-plasma within an effective 
kinetic theory approach. In this way, the compatibility of the employed 
quasiparticle ansatz with the apparent low viscosities of the strongly 
coupled deconfined gluonic medium is shown. 
\end{abstract} 

\end{frontmatter} 



\section{Introduction\label{sec.1}}

Within the past years, an immense effort has been put into revealing 
the nature and properties of deconfined strongly interacting matter both 
theoretically and experimentally in relativistic 
heavy-ion collisions. The aim is to understand characteristics of QCD 
matter such as its equation of state (EoS), collective behaviour or transport 
properties which have a wide range of implications in cosmology and 
astrophysics. While the equation of state 
describes the system in thermal equilibrium, the transport coefficients, 
such as viscosities, characterize the system's ability to relax from 
nonequilibrium towards equilibrium. 

The success of perfect fluid hydrodynamic calculations in describing 
the collective flow observed in heavy-ion collisions~\cite{Teaney,Kolb,Huovinen} 
suggests small viscosities of the created medium which has 
also lead to viewing it as being strongly coupled~\cite{Gyulassy}. Nonetheless, 
in particular the shear viscosity cannot be arbitrarily small due to 
unitarity~\cite{Starinets}. Besides, in weak coupling certain 
relations hold among different transport coefficients, 
cf.~e.~g.~\cite{Majumder}. 

Apart from the first-principle calculations of bulk ($\zeta$) and shear ($\eta$) 
viscosities by means of lattice QCD~\cite{Meyer1,Nakamura} a 
variety of other approaches was proposed. Among these the rigorous 
perturbative calculations starting either from the Boltzmann 
equation~\cite{Arnold1,Arnold2} or from the Kubo 
formalism~\cite{Jeon} have to be mentioned. Besides, considerations 
employing different spectral functions~\cite{PeshierCassingKharzeev} 
or numerical transport calculations of the Boltzmann equation~\cite{Xu} 
have been made. 

Here, we adress viscosities by viewing the gluon-plasma as 
composed of quasiparticle excitations. 
The underlying quasiparticle model (QPM) was successfully 
tested to describe lattice QCD results of the EoS~\cite{Bluhm1,Bluhm2}. 
In the QPM, gluon and quark quasiparticles obey 
dispersion relations, where the entering self-energies 
$\Pi$, in general, depend on temperature $T$ and chemical potential $\mu$ 
both explicitly and also implicitly 
via a phenomenological effective coupling $G^2(T,\mu)$~\cite{Peshier}. 
We extend this picture to nonequilibrium 
systems by means of an effective kinetic theory for gluon 
quasiparticles, i.~e.~$\mu=0$ in the following. 

\section{Effective kinetic theory for quasiparticle excitations\label{sec.2}}

For a system in nonequilibrium, the gluon quasiparticle dispersion relation no 
longer depends on an uniquely defined temperature but becomes space-time 
dependent, $E(x)=\sqrt{\vec{p}^{\,2} + \Pi(x)}$. In this case, 
the space-time dependence of the distribution function $b(x,p)$ 
is governed by the Boltzmann equation 
\begin{equation}
 \label{equ:1}
 \left(p^\alpha (x)\partial_\alpha + \sqrt{\Pi(x)} F^\alpha(x) 
 \frac{\partial}{\partial p^\alpha(x)}
 \right) b(x,p) \equiv \mathcal{D} b(x,p) = \mathcal{C}[b(x,p)] \,,
\end{equation}
where $\mathcal{C}[b(x,p)]$ is the collision term. The force 
$F^\alpha = \partial^\alpha \Pi /(2\sqrt{\Pi})$ 
satisfies $p_\alpha F^\alpha =0$ such that the 
spatial gradient of the self-energy acts as an external force changing the 
momenta of the quasiparticles between collisions. 
The scattering interaction conserves locally energy and 
momentum which results in a vanishing collision term when multiplied by 
$p^\nu$ and integrated over three-momentum $\vec{p}$. Hence, 
an energy-momentum tensor 
\begin{equation}
 \label{equ:2}
 T^{\mu\nu}(x) = d \int \frac{d^{\,3} \vec{p}}{(2\pi)^3 E(x)} p^\mu(x) p^\nu(x) b(x,p) 
 + g^{\mu\nu} B(\Pi(x))
\end{equation}
can be defined which, as a consequence of the Boltzmann equation, obeys 
energy-momentum conservation $\partial_\mu T^{\mu\nu}(x) = 0$ under the condition 
\begin{equation}
 \label{equ:3}
 \frac{\partial B}{\partial\Pi(x)} = -\frac12 q(x) \,,\,\, 
 q(x) = d \int \frac{d^{\,3} \vec{p}}{(2\pi)^3 E(x)} b(x,p) \,,
\end{equation}
for the mean field $B$. 
Here, $q(x)$ is an auxiliary field~\cite{Jeon}, 
$g^{\mu\nu}=diag(1,-1,-1,-1)$ and $d$ is the 
number of degrees of freedom. As a result of Eq.~(\ref{equ:3}), 
the space-time dependence of 
the self-energy is determined by the auxiliary field. Thus, the Liouville 
operator $\mathcal{D}$ on the left hand side of Eq.~(\ref{equ:1}) is a functional 
of the distribution function $b(x,p)$ itself. 

In thermal equilibrium, characterized by a local distribution function 
$b^0(x,p)=(e^{p^\alpha u_\alpha /T} - 1)^{-1}$, where the fluid four-velocity 
$u_\alpha$ satisfies $u_\alpha u^\alpha =1$, 
one has to demand that $\left.\Pi(q(x))\right|_{f^0}\equiv \Pi(T)$ to 
recover equilibrium results from the energy-momentum tensor~(\ref{equ:2}). 
In fact, by comparing Eq.~(\ref{equ:2}) evaluated for 
$b^0(x,p)$ in the local rest frame with 
$T^{\mu\nu}_{(0)}=\epsilon u^\mu u^\nu - P (g^{\mu\nu}-u^\mu u^\nu)$, 
the quasiparticle model expressions 
for energy density $\epsilon$ and pressure $P$ are recovered, 
cf.~\cite{Bluhm1,Peshier}. Moreover, Eq.~(\ref{equ:3}) represents the 
nonequilibrium generalization of the stationarity condition ensuring 
thermodynamic self-consistency and 
also implying that the principles of statistical mechanics assure the 
physical meaning of $b^0(x,p)$~\cite{Gorenstein}. In this way, $T^{\mu\nu}(x)$ 
in Eq.~(\ref{equ:2}) represents the general form 
for an isotropic fluid composed of quasiparticle excitations including only 
$g^{\mu\nu}$ and $u^\mu$ which, in addition, satisfies thermodynamic 
self-consistency in equilibrium. 

\section{Bulk and shear viscosities\label{sec.3}}

The calculation of transport coefficients from $T^{\mu\nu}(x)$ as the first-order 
corrections to thermal equilibrium is straightforward. Assuming small deviations 
from equilibrium, $b(x,p)=b^0(x,p)+\delta b(x,p)$ with $\delta b\ll b^0$, 
$T^{\mu\nu}(x)$ can be decomposed into 
$T^{\mu\nu}=T^{\mu\nu}_{(0)}[b^0]+\delta T^{\mu\nu}[\delta b]$. Expanding 
$B(\Pi(q(x)))$ in terms of small deviations from its equilibrium value 
and approximating $E(x)$ to lowest order by $E=\sqrt{\vec{p}^{\,2}+\Pi(T)}$, 
$\delta T^{\mu\nu}$ to lowest order in $\delta b$ can be written as 
\begin{equation}
\label{equ:4}
 \delta T^{\mu\nu} = d \int \frac{d^{\,3} \vec{p}}{(2\pi)^3 E} \delta b(x,p) 
 \left(p^\mu p^\nu - \frac12 g^{\mu\nu} \left[ q \frac{\partial \Pi}{\partial q}
 \right]_{b^0}\right) \,.
\end{equation}

Assuming that collisions always result in an 
exponentially fast restoration of local equilibrium with the relaxation time 
$\tau$, all the complexity of $\mathcal{C}[b]$ is encoded in 
$\tau$~\cite{Reif}. Correspondingly, $\delta b$ can be approximated by 
$\delta b = -\mathcal{C}[b]/\tau$, which in turn can be expressed by the 
Boltzmann equation~(\ref{equ:1}). Thus $\delta b$ in Eq.~(\ref{equ:4}) is a 
functional of $\mathcal{D}$ acting on $b$ reading to lowest order 
\begin{equation}
\label{equ:5}
 \delta b(x,p) = -\frac{\tau}{E} \left(p^\alpha \partial_\alpha 
 - \frac12 \vec{\nabla} \Pi(T) \frac{\partial}{\partial\vec{p}}\right) b^0(x,p) \,.
\end{equation}

Considering gluonic quasiparticles with $\Pi(T)=\frac12 T^2 G^2(T)$, we note first that 
for recovering the equilibrium QPM one needs to identify 
$\Pi(q)=\tilde{q}/(\beta_0 \log [\lambda\{\sqrt{\tilde{q}}-T_s\}/T_c]^2)$, 
where $\left.\tilde{q}\right|_{b^0}\equiv \left.(\mathcal{N}q)\right|_{b^0}\equiv T^2$ 
and $\beta_0 = 11/(8\pi^2)$. Furthermore, one finds 
$q\left.(\partial\Pi /\partial q)\right|_{b^0}\equiv T^2(\partial\Pi(T)/\partial T)$. 
When evaluating Eq.~(\ref{equ:4}) supplemented by Eq.~(\ref{equ:5}), it is 
appropriate to replace the convective derivatives of $T$ and $\vec{u}$ by spatial 
gradients. In line with the Chapman-Enskog strategy, 
the space-time dependence of $b$ is assumed to be determined only by $T$ and $u^\mu$, 
constituting $b^0$, as well as their gradients. Then, the conservation equations are 
given in first approximation by the equations of motion and energy of an ideal 
fluid. To make the decomposition of $b(x,p)$ unique, 
we employ the Landau-Lifshitz condition~\cite{Landau}, 
i.~e.~$u_\nu T^{\mu\nu}\equiv u_\nu T^{\mu\nu}_{(0)}=\epsilon u^\mu$, 
implying $\delta T^{00}=0$ in the local rest frame. 

For the considered gluonic system, the bulk and 
shear viscosities are the only independent transport coefficients characterizing 
the fluid. In case of small deviations from equilibrium, they are obtained 
from the spatial part of the nonequilibrium energy-momentum tensor 
$\delta T^{ij} = - \zeta \delta^{ij} \partial_k u^k + \eta W^{ij}$ as 
coefficients of the scalar and the traceless part 
$W^{ij}=\partial^i u^j +\partial^j u^i -\frac23 g^{ij} \partial_k u^k$. In 
the local rest frame one finds 
\begin{eqnarray}
\label{equ:7}
 \zeta & = & \frac{d}{3T} \int \frac{d^{\,3} \vec{p}}{(2\pi)^3 E} \tau 
 b^0(1+b^0)\left(\frac{\vec{p}^{\,2}}{3E}-\left[E-T\frac{\partial E}{\partial T}\right]
 \frac{\partial P}{\partial\epsilon} \right) 
 \left\{2 T^2 \frac{\partial\Pi(T)}{\partial T^2}-\Pi(T) \right\} \,,\\ 
\label{equ:8}
 \eta & = & \frac{d}{15T} \int \frac{d^{\,3} \vec{p}}{(2\pi)^3 E} \tau 
 b^0(1+b^0) \frac{\vec{p}^{\,4}}{E} \,.
\end{eqnarray}
From Eqs.~(\ref{equ:7}) and~(\ref{equ:8}) it is clear, that only $\zeta$ is 
significantly influenced by the medium-dependent quasiparticle dispersion 
relations $E$, cf.~also~\cite{Redlich}, while in $\eta$ the mean field 
contributions vanish. 

\begin{figure}[t]
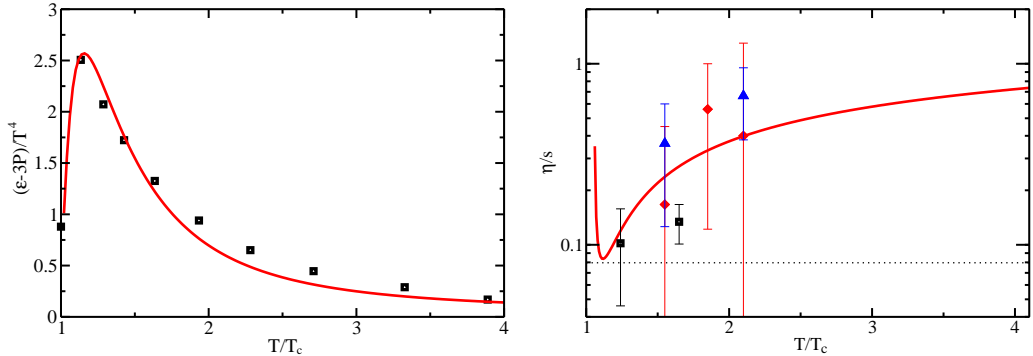

\centering
\includegraphics[scale=0.27]{EoS4.eps}	       
\hspace{2mm}
\includegraphics[scale=0.27]{etadivs5.eps}	       
\caption[]{Left: Comparison of QPM results for the scaled 
interaction measure with lattice QCD results for pure SU(3)~\cite{Boyd} (boxes). 
The adjusted model parameters read $d=16$, $T_s=0.53\,T_c$, $\lambda=2.88$, where in line 
with lattice QCD results we set $T_c=271$ MeV. Right: Corresponding QPM result 
for $\eta /s$, employing $a_\eta =6.8$ in the ansatz for the relaxation time 
$\tau$, compared with lattice QCD results (boxes from~\cite{Meyer1}, 
diamonds and triangles from~\cite{Nakamura}). 
In addition, the unitarity limit~\cite{Starinets} $\eta /s=1/(4\pi)$ (dotted 
curve) is depicted.}
\label{figure1}
\end{figure}

To quantify $\eta$, for instance, the parameters of 
the effective coupling, $T_s$ and $\lambda/T_c$, 
are first adjusted to lattice QCD results of the EoS, for example, of the scaled 
interaction measure $(\epsilon - 3P)/T^4$ (left panel of 
Fig.~\ref{figure1}). 
In addition, for $\tau$ required in Eq.~(\ref{equ:8}), 
we employ an ansatz inspired by previous work~\cite{Hosoya}, 
$\tau^{-1}=a_\eta/(32\pi^2) T G^4 \log (a_\eta\pi / G^2)$, 
where the QCD running coupling 
is replaced by our effective coupling $G^2(T)$. The corresponding result for the 
shear viscosity to entropy density ratio $\eta /s$ is depicted in 
Fig.~\ref{figure1} (right panel). 

\section{Conclusion \label{sec.4}}

We discuss the bulk and shear viscosities by means of an effective kinetic theory 
for gluon quasiparticle excitations in the relaxation time approximation. Thermodynamic 
self-consistency necessary in such a quasiparticle description turns out to be 
solely a consequence of energy-momentum conservation. Our numerical results for 
$\eta /s$ are in agreement with available lattice QCD results~\cite{Meyer1,Nakamura} 
and with other approaches~\cite{PeshierCassingKharzeev}. Moreover, $\eta /s$ 
exhibits the expected minimum close to $T_c$ 
similar to classical fluids~\cite{Csernai}, which here is mostly driven 
by $\tau$. Contrary to the popular view, 
where a large quasiparticle mean free path implies 
large $\eta /s$~\cite{LindenLevy}, our results suggest that a quasiparticle 
description is still admitted for the strongly coupled gluon-plasma. 

\vspace{2mm}
\noindent
The authors thank A.~Peshier, C.~Sasaki and R.~Schulze for valuable discussions. 
The work is supported by 06 DR 136, GSI-FE and the Polish Ministry of Science. 

\end{document}